\documentclass[%
 reprint,
 amsmath,amssymb,
 aps,
]{revtex4-1}

\usepackage[utf8]{inputenc}
\usepackage{graphicx}
\usepackage{dcolumn}
\usepackage{bm}

\begin{document}

\preprint{APS/123-QED}

\title{Quark-antiquark potentials in non-perturbative models}
\thanks{Poster presented at the XIV Hadron Physics, $18^{th} - 23^{nd}$ March 2018, in Florian\'opolis (Brazil).}%

\author{C. Mena}
 \email{cstiven.mc2@gmail.com}
\author{L. F. Palhares}%
 \email{leticia.palhares@uerj.br}
\affiliation{%
 UERJ -- Universidade do Estado do Rio de Janeiro,\\
 Departamento de F\'\i sica Te\' orica, \\
Rua S\~ao Francisco Xavier 524, 20550-013, Maracan\~a, Rio de Janeiro, Brasil
}%




\begin{abstract}

In these proceedings, we investigate non-perturbative models for Quantum Chromodynamics (QCD) motivated by the behavior of the Landau-gauge lattice gluon propagator with the purpose of testing their validity in the perturbative regime of strong interactions and to explore their behavior in the infrared region. In particular, we discuss the potentials between heavy quarks and antiquarks, since this observable might reveal the appearance of confining properties in these non-perturbative models through a linear growth at large and intermediate distances. 
\begin{description}

\item[Keywords]
Quantum Chromodynamics, Confinement and Effective Models for QCD, Static Potentials.
\end{description}
\end{abstract}
\pacs{Valid PACS appear here}
\maketitle

\section{\label{sec:level1}Introduction}

Quantum chromodynamics, the theory that allows us to describe the fundamental interaction between quarks and gluons, is asymptotically free for high energies. This is experimentally verified in scattering processes, such as Deep Inelastic Scattering, in which perturbative predictions with quasi-free quarks explain well the data \cite{Soding1981,Patrignani:2016xqp}.

However, due to the absence of free, asymptotic states of quarks or gluons in the measure spectra and the exclusive observation of colorless bound states (hadrons), the phenomenon of color confinement has been conjecture as an essential feature of Strong Interactions. Even though there is still no analytical proof that QCD produces confinement, there are numerical results from Lattice QCD simulations that indicate confining properties \cite{Bali:1998de,Brambilla:1999ja,Burnier:2014ssa}.

The extreme difficulty in describing confinement from first principles in QCD is linked to the fact that the strong coupling constant shows a very marked behavior in relation to the energy in which the processes are studied. As the energy decreases the coupling constant grows in such a way that perturbation theory fails. Therefore one must resort to other techniques to address the problem. There are phenomenological and non-perturbative models that attempt to explain color confinement, and  -- even if a fully successful description is still unavailable -- they allow us to address several aspects of QCD. Some attempts to understand confinement point to the interaction potential between a heavy quark-antiquark pair, which should grow linearly with the separation scale between them, guaranteeing thus that they will stay confined within meson bound states. This property is well-known to be connected with an area law for the Wilson loop in pure gauge theory and has been explored in  different contexts, receiving robust corroboration from Lattice QCD investigations \cite{Greensite:2011zz}. 

A completely different approach to try to understand confinement is to build a consistent nonperturbative theory that is simultaneously valid in the infrared regime of Strong Interactions and amenable to (semi-)analytic investigations. Following this line, models based on taking into account the presence of Gribov gauge ambiguities \cite{Gribov1977} in the gluon path integral have been proposed.
The main analytical technique of quantizing a Gauge Field Theory in the continuum requires gauge fixing, because in the functional integration the gauge fields present a redundancy of equivalent physical states due to the local symmetry. In QCD this is usually done through the proposal of Faddeev and Popov \cite{Faddeev:1967fc}, which is consistent at the perturbative level. As shown by Gribov \cite{Gribov1977}, equivalent fields remain being integrated. Gribov's proposal was then to restrict the field integration to a region called the Gribov Region \footnote{Although in this region equivalent configurations of the fields are still present \cite{VANBAAL1992259}}. Zwanziger was able to formulate a local action that implements the Gribov idea by modifying the theory by a so-called horizon condition  \cite{zwanziger1989local}, which sets the value of a new parameter with unit of mass. In fact, in the seminal work of V. Gribov a massive parameter is already present, being known as the Gribov mass.  In the Gribov-Zwanziger framework, the gluon propagator is modified by the presence of the horizon, presenting two poles with complex-conjugated masses. This pole structure violates the axiom of reflection positivity and hinders the asymptotic-particle interpretation for the gluon; a feature consistent with gluon confinement. The search of other confining properties in these non-perturbative models is of course of great importance to establish their role as consistent infrared realizations of QCD.

With this in mind, we explore the interaction potential between heavy quarks and antiquarks for different non-perturbative models that modify the gluon propagator in the infrared \footnote{Other recent modelings and applications of quark-antiquark potentials can be found e.g. in Refs. \cite{Gonzalez:2017izt,Cucchieri:2017icl}}. Our aim is to investigate the regime of validity of these descriptions and whether such models can produce quark confinement through a linearly rising potential.
\newpage

\section{\label{sec:level2}Confinement and the \\
Quark -- Antiquark Static Potential}

%
The phenomenological string model \cite{flux-tube:1979,flux-tube:1984} provides a qualitative view of confinement in which the potential between a heavy quark-antiquark pair grows linearly when we try to separate them, confining them to bound states.  

\begin{figure}[!ht]
\centering
\includegraphics[scale=0.31]{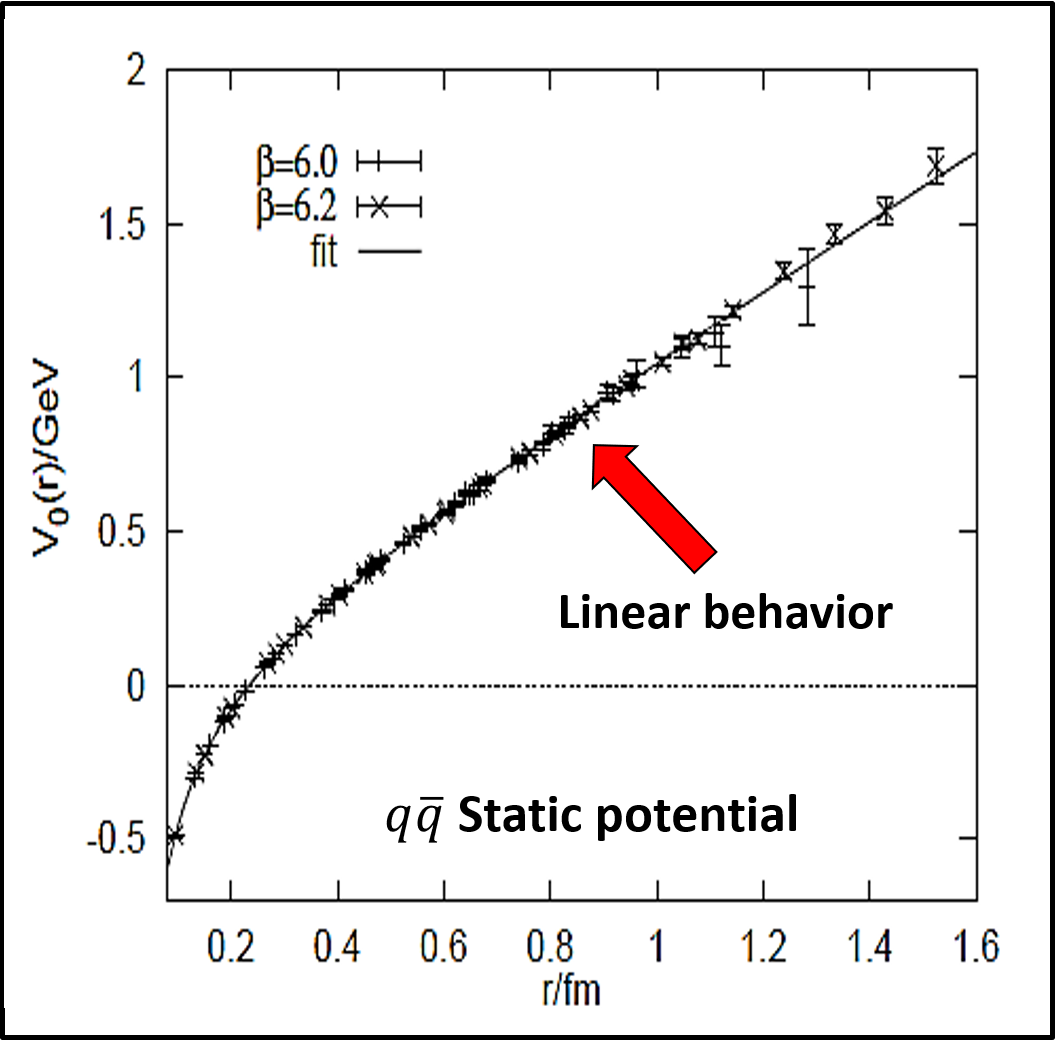}
\caption{$q$-$\overline{q}$ quenched potential calculated in Lattice QCD \cite{Brambilla:1999ja}.} 
\label{sbmlattice}
\end{figure}

In Fig.\,\ref{sbmlattice}, we show results from Lattice QCD in the quenched approximation \cite{Brambilla:1999ja}, valid for infinitely massive quarks. At very short distances, asymptotic freedom prevails and the quark-antiquark potential resembles a Coulomb-like potential.
For intermediate and large distances the potential behavior departs significantly from the perturbative, Coulomb form. In the string model, this is supposedly due to the formation of a color flux-tube, in which a color-electric field connects the $q$-$\overline{q}$ (static) pair thereby forming a kind of string with a nearly-constant cross-sectional area.

According to this model, the energy stored in the flux-tube increases linearly with the separation distance between the static $q$-$\overline{q}$ pair, so one would need an infinite energy to separate the pair at an infinite distance, i.e. to produce free quarks and antiquarks.
This is not the case in full QCD. As the quark-antiquark pair is separated, and due to the energy stored in the flux-tube, the creation of a new pair becomes more energetically favorable, as the light fermions that can be excited from the quantum vacuum. This results in the rupture of the flux-tube ("String Breaking") and the potential tends to saturate.
In the current analysis, however, we shall concentrate ourselves in the heavy quark approximation and look for the linearly rising potential as a sign of confinement.
%
%
%
%
\section{\label{sec:level3}Non-Perturbative Effective Models}
\begin{figure}[!ht]
\centering
 \includegraphics[scale=0.45]{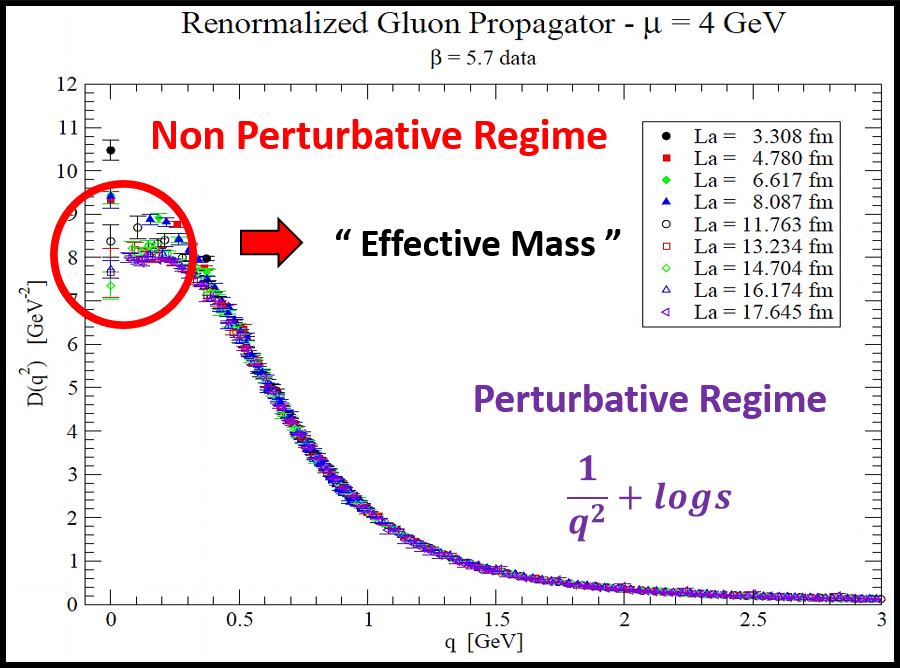}
 \caption{ Gluon propagator calculated in Lattice QCD \cite{Oliveira:2012eh}.} 
 \label{pgtrlattice}
 \end{figure}

%
%
QCD is a gauge theory, and it is exactly gauge invariance that tells us that the gluon must be a massless particle. Even so, as can be seen in Fig.\,\ref{pgtrlattice}, Landau gauge Lattice QCD simulations show the emergence in the infrared region of a finite value for the gluon propagator at zero momentum. This may be interpreted as the generation of an effective mass which disappears in the perturbative limit.

Here, in order to search for confinement features in the heavy $q$-$\overline{q}$ potential, we will use different models motivated by the behavior of the lattice gluon propagator, namely:
The Massive Gluon Model \cite{MANDULA1987117,Oliveira:2010xc}, The Gribov-Zwanziger Model (GZ) \cite{zwanziger1989local} and the Refined Gribov-Zwanziger Model (RGZ) \cite{PhysRevD.78.065047}. These will be compared to the Perturbative approach at short distances and Lattice QCD  in the IR regime.
 \begin{figure}[!ht]
\centering
 \includegraphics[scale=0.25]{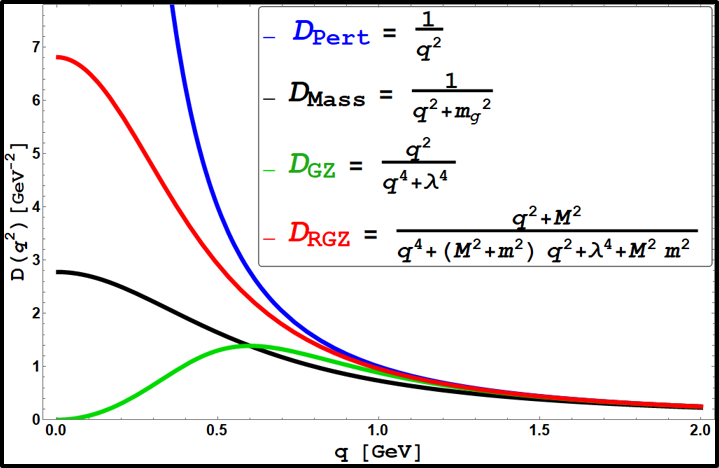}
 \caption{Scalar form factor of the gluon propagator (c.f eq.(\ref{Propag})) for the different models as a function of the momentum.} 
 \label{pgtr}
 \end{figure}

In Fig.\,\ref{pgtr}, we plot the momentum dependence of the scalar function in the gluon propagator for the different analytical models.
The massive gluon model adopts a fixed gluon effective mass ($m_{g}^{2}$) that is assumed to come from the strong interaction in the IR regime. The behavior of the propagator in this model is qualitatively similar to Lattice QCD one, in Fig.\,\ref{pgtrlattice}.
The Gribov-Zwanziger (GZ) model, shows a Gribov mass ($\lambda$) term -- coming from the horizon condition -- that modifies the IR behavior of the gluon propagator which tends to zero as the momentum vanishes, being incompatible with the lattice results \cite{Oliveira:2012eh} in the deep IR.

Since the GZ model does not reproduce the Lattice QCD propagator, a refinement with the inclusion of condensates, $M^{2}$ and $m^2$, was proposed \cite{PhysRevD.78.065047}. In the RGZ framework, $M^{2}$ ensures that the gluon propagator is nonzero at zero momentum, while $m^2$ is indispensable to find a good quantitative agreement with the Lattice data.

\noindent
The different models will be discussed in more detail in a paper \cite{Mena:Plrs}.
\section{\label{sec:level4}Calculating the $q$-$\overline{q}$ Potential}
\begin{figure}[!ht]
\centering
\includegraphics[scale=0.22]{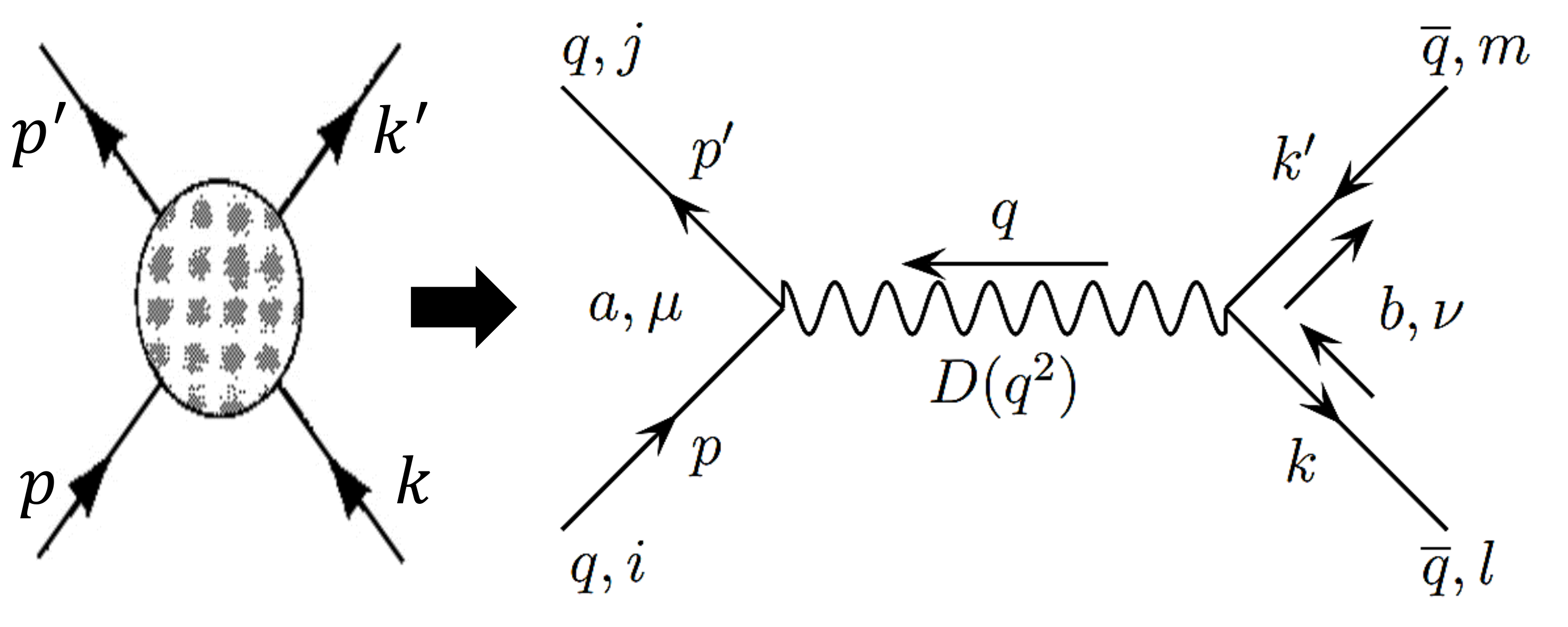}
 \caption{Leading--order contribution to the quark-antiquark interaction process.} 
\label{qaqprocess}
\end{figure}

Supposing heavy quarks, a single diagram contributes in the tree-level approximation of the $q$-$\overline{q}$ interaction (Fig.\,\,\ref{qaqprocess}), that can be represented by the matrix element $\mathcal{M}$ and written like \cite{Peskin1995}:
\begin{eqnarray}
\label{Mqaq}  
i\mathcal{M}= - \mathcal{\overline{U}}_{(p')}^{s'}
[ig \gamma^{\mu} t^{a}_{ji}] \,
\mathcal{U}_{(p)}^{s}
[ \mathcal{D}_{\mu\nu}^{ab}(q^{2})]
\mathcal{\overline{V}}_{(k)}^{r} [ig \gamma^{\nu} t^{b}_{lm}] \mathcal{V}_{(k')}^{r'}\, , \hspace{16pt}
\end{eqnarray}
where \,$\mathcal{U},\mathcal{V}$ and $\mathcal{\overline{U}},\mathcal{\overline{V}}$ are the quark and antiquark spinors, i.e. solutions of the Dirac equation, as a function of their four-momenta, respectively.\,\,The factors $g$, $\gamma^{\mu}$ and $t^{a}$ are the strong coupling constant, the Dirac matrices and the $SU(3)$ group generators, respectively. $a, b \equiv\,1,..,8$ and $\mu, \nu \equiv\,0,..,3$ are color and Lorentz indices, respectively. $\mathcal{D}_{\mu\nu}^{ab}(q^{2})$ is the gluon propagator:
\begin{eqnarray}
\mathcal{D}_{\mu\nu}^{ab}(q^{2}) = -i \, \delta^{ab} \,  \mathcal{D}(q^2) \, \big( \, g_{\mu\nu}-(1-\xi )\frac{q_{\mu}q_{\nu}}{q^{2}} \, \big) \, ,
\label{Propag}
\end{eqnarray}

\noindent
where $\xi$ is the gauge parameter.

We  can simplify expression (\ref{Mqaq}), so that it is explicitly independent of the gauge parameter, yielding:
\begin{eqnarray}
i\mathcal{M} =  -\, 4\pi i \,\alpha_{s}  \,\mathcal{C}_{F}\,\mathcal{D}(q^{2}) \, [\mathcal{\bar{U}}_{p'}^{s'}\,\gamma^{\mu}\, \mathcal{U}_{p}^{s}] \, [\mathcal{\bar{V}}_{k}^{r}\, \gamma_{\mu} \, \mathcal{V}_{k'}^{r'}]\,,
\end{eqnarray}

\noindent
with $C_{F}$ being a  color factor.

The amplitude $\mathcal{M}$ may be connected to the one obtained through Quantum Mechanics. This will allow us to extract an associated potential.
We have that the scattering amplitude for the initial $|\,i \, \rangle$ and final state $\langle \, f\,|$ of two particles in non-relativistic quantum mechanics can be expressed as:
\begin{equation}
\label{Ampfi}
\langle \,f\,|\,i\, \rangle \,=\,i\, \mathcal{M} \,(2\, \pi)^{4} \, \delta^{4}(p+k-p'-k') \, ,
\end{equation}
where $p,p',k$ and $k'$ are the four-momenta of the particles in this process. This same quantity can be expressed in Quantum Mechanics in the Born approximation by:
\begin{equation}
\label{fiPotencial}
\langle \,f\,|\,i\, \rangle \,=\,-\,i\,V(\overrightarrow{p}'-\overrightarrow{p}) \, (2 \pi)^{4} \delta^{4}(\, p + k - p' - k' \,) \; ,
\end{equation}

\noindent
where a potential $V$ is associated to the interaction.

So, one may use the matrix element $\mathcal{M}$ calculated from the Quantum Field Theory for the quark-antiquark process in the non-relativistic approximation to obtain a potential associated with the interaction process.
%
%
\section{\label{sec:level5}RESULTS and remarks}
%
Finally, using these approximations, we will calculate the leading-order quark-antiquark potentials as functions of the separation distance $r$ between the particles for the perturbative case and the massive, GZ and RGZ models.\,\,The computation of the potentials yields:
\begin{align}
\label{vp}
& V_{Pert}=-\alpha_{s}\,C_{F}/r \, , \\
\label{vmass}
& V_{Mass}=-\alpha_{s}\,C_{F}\, e^{-m_{g}r} /r \, , \\
\label{vgz}
& V_{GZ}=-\alpha_{s}\,C_{F}\, e^{-\lambda r/ \sqrt{2} }\, \cos{(\lambda r/ \sqrt{2} )}/r \, , \\ 
\label{vrgz}
& V_{RGZ}=-\alpha_{s}\,C_{F}\, e^{- a\,b\,r} \left( \cos{(ac\, r)}  + k \, \sin(ac\, r) \right)/r \, ,
\end{align}
where $a,b,c$ and $k$ are functions of the RGZ model parameters $\lambda$, $M$, $m$ and  $\alpha_{s}=g^{2}/4\pi$.

 \begin{figure}[!ht]
\centering
 \includegraphics[scale=0.24]{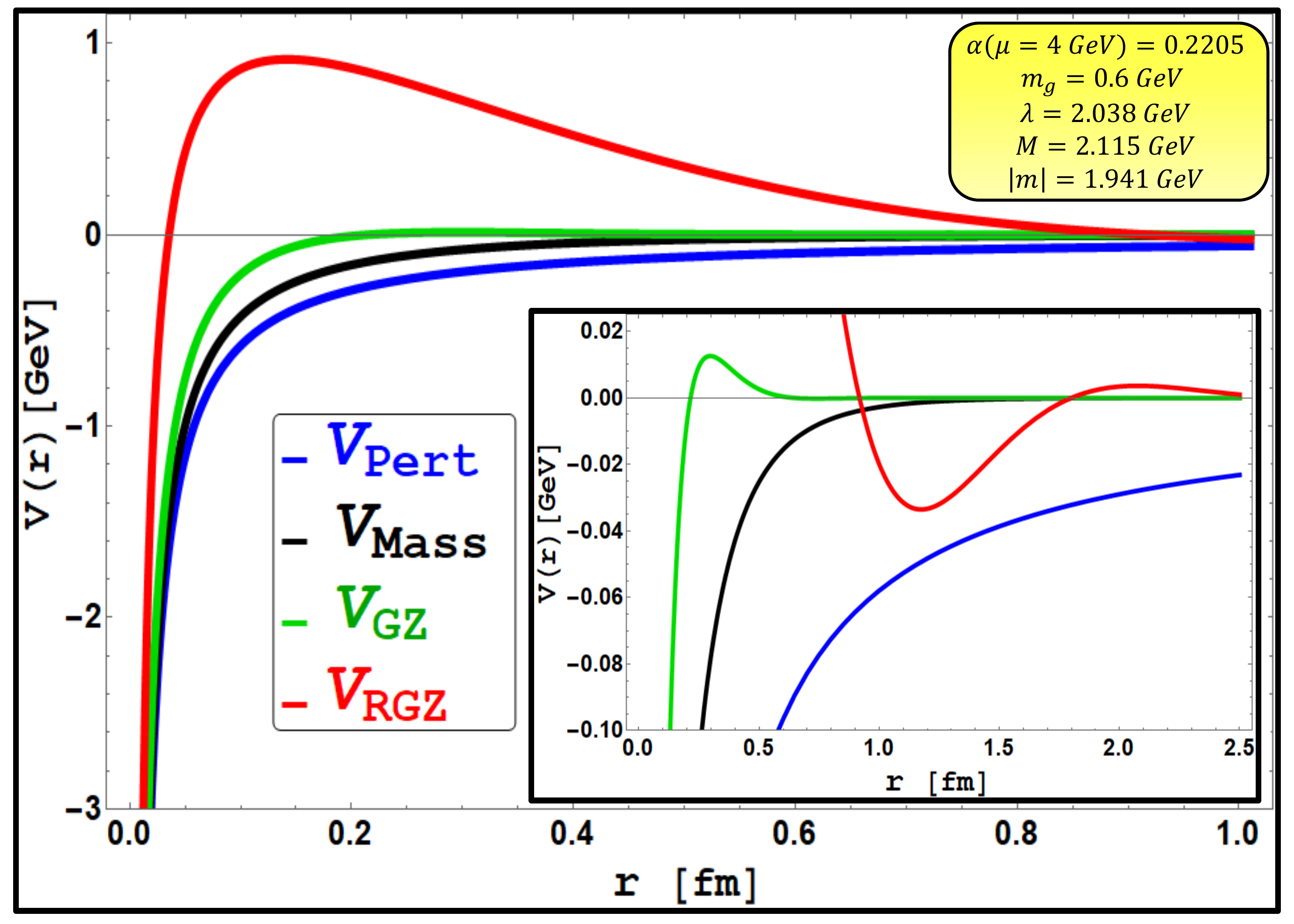}
 \caption{Potentials for the different models as functions of the distance.\,\,In the upper right corner the values of the parameters used are displayed.} 
  \label{cpgraph}
 \end{figure}
 
We can recognize that the perturbative potential, eq.\,(\ref{vp}), has a Coulomb-like behavior, analogous to electromagnetism with the difference that an extra color factor and the strong coupling constant appear.\,\,In the massive potential, eq.\,(\ref{vmass}), the typical exponential decay that depends on the mass of the particle emerges, while in the GZ potential, eq.\,(\ref{vgz}), in addition to the Yukawa decay, an oscillatory term arises.\,\,The RGZ potential, eq.\,(\ref{vrgz}), shows a mass-like behavior too, but includes a combination of oscillating terms.\,\,All these characteristics will be reflected in the potentials' plots, displayed in Fig.\,\ref{cpgraph}.
%

In Fig.\,\ref{cpgraph}, we can see that the non-perturbative models reduce to the perturbative potential for short distances, $ r \lesssim 0.02 $ fm, with the RGZ model requiring the shortest distance to reproduce the perturbative result. This feature is very important and reassuring, since it shows that these models are consistent with the well-tested perturbative approach of high-energy QCD.

For the intermediate-distance regime, we can appreciate that the Massive and GZ models have similar behaviors, and they tend to zero faster than the Perturbative potential due to the presence of an effective infrared mass for the gluon. Such a faster damping of the potential is  expected for short-range interactions like QCD. As one increases the separation scale, the RGZ model also displays a potential that rapidly approaches zero, but shows a very distinct behavior: it has a rapid growth upto $r \sim 0.12$ fm, with the most important feature being the sign change\footnote{A similar sign change occurs in the GZ model, but the effect is much less pronounced.}.
In this region, the behavior of the RGZ potential could be an indication of confinement, since to separate the quark--antiquark pair one would need more and more energy to do it.\,\,Despite this promising result in the intermediate region, in this leading-order approximation, the RGZ potential at a point close to $ r\sim 0.12 $ fm decreases again, entering an oscillating regime. 
It is possible that this feature is an artifact of the approximations adopted, so that a more thorough study of the effects of systematically including higher-order interactions is welcome to resolve this issue. This investigation is currently underway and we hope to report more findings soon \cite{Mena:Plrs}.
%
%
\section{\label{sec:level6} SUMMARY AND outlook}

Color confinement is still an open problem in QCD. There is a plethora of effective models that address this issue.
In particular, the string model provides a simple qualitative view of the binding of heavy quark-antiquark states, mesons, and of confinement. It is reflected in the behavior of the associated $q-\overline{q}$ potential  which has a dominant Coulomb-like part at short distances, while for large distances -- that is, the non-perturbative regime of QCD -- displays a linear growth, which can be considered a sign of confinement.

Ideally, other approaches to the confining QCD regime should be able to reproduce a linearly rising potential between heavy quarks and antiquarks. 
Here, we started searching for this feature in different models that include non-perturbative characteristics such as effective masses and variations in the propagators of the theory, motivated by Landau-gauge lattice results and based on the modifications brought about by the presence of Gribov gauge ambiguities in the path integral.

We have calculated the leading-order potentials and verified that their behavior is compatible with the perturbative approach for short distances, while displaying nontrivial features for intermediate and large separation scales. Further investigation is underway \cite{Mena:Plrs} to establish whether these models are able to produce a linearly rising potential for at least an intermediate distance region that broadens as more interactions are included.\\

\vspace{-.8cm}
\acknowledgments
%
C. Mena would like to thank the organizers for the rich discussion environment at Hadron Physics 2018, where this poster was presented, and CAPES for Master and PhD fellowships. This work was partially supported by the Brazilian agencies CNPq, CAPES and FAPERJ.

\bibliographystyle{apsrev4-1} 
\bibliography{refs}

\end{document}